\documentclass[journal]{IEEEtran}
\usepackage[utf8]{inputenc}
\usepackage{graphicx}
\usepackage{enumerate}
\usepackage{setspace}
\usepackage{subfigure}
\usepackage{url}
\usepackage[cmintegrals]{newtxmath}
\usepackage[linesnumbered, ruled, vlined]{algorithm2e}%[ruled,vlined]{
\usepackage{balance}

\usepackage{algpseudocode}
\SetAlFnt{\footnotesize}
\SetAlCapFnt{\footnotesize}
\SetAlCapNameFnt{\footnotesize}
\SetInd{0.1cm}{0.35cm}

\usepackage{adjustbox}
\usepackage{units}
\usepackage{upquote}
\usepackage{varwidth,array,ragged2e}
\usepackage{tikz}

\begin{document}

\title{MPTCP Linux Kernel Congestion Controls}
%\title{Coupled Congestion Control Algorithms \\ of MPTCP Linux Kernel Implementation}

\author{Bruno~Yuji~Lino~Kimura, and Antonio~Alfredo~Frederico~Loureiro
\thanks{B.~Y.~L.~Kimura is with the Institute of Science and Technology of the Federal University of São Paulo (ICT/UNIFESP), São José dos Campos -- SP, Brazil. Email: bruno.kimura@unifesp.br}
\thanks{A.~A.~F.~Loureiro is with the Department of Computer Science of the Federal University of Minas Gerais (DCC/UFMG), Belo Horizonte -- MG, Brazil. Email: loureiro@dcc.ufmg.br}
%\thanks{Manuscript received Month Day, 2017; revised Month Day, 2017.}
}

\markboth{Technical Report}{MPTCP Linux Kernel Congestion Controls}

\maketitle

\begin{abstract}
MultiPath TCP (MPTCP) is a promising protocol which brings new light to the TCP/IP protocol stack ossification problem by means of an impactful innovation of the transport layer. A MPTCP connection consists of a set of one or more subflows, where each subflow offers an alternative path to reach a target remote end-system. However, simply applying the standard TCP congestion control on each subflow would give an unfair resource sharing. Various subflows of a connection would dispute bottleneck links with regular single-path TCP connections, leading them to starvation conditions. To deal with this concern, a multipath congestion control algorithm adjusts the sending operation of all subflows in a coupled fashion in order to achieve various objectives, e.g., friendliness, responsiveness,  throughput improvement, and congestion balance. In this report, we describe the four coupled congestion control algorithms deployed in the MPTCP Linux kernel implementation, namely: LIA, OLIA, BALIA, and wVegas. We provide a concise material with technical details of each algorithm, while summarizing all of them together from a single notation.
\end{abstract}


\section{Introduction}

MultiPath TCP (MPTCP)~\cite{rfc6824} is a set of TCP extensions that enable reliable data transfer over multiple end-to-end paths between source and destination nodes. Two benefits are forthwith achieved~\cite{rfc6824}: it improves end-to-end throughput, while allowing greater tolerance to communication failures. As a general-purpose protocol, MPTCP has a broad range of potential uses. Enabling efficient and resilient mobile communications~\cite{Paasch.et.al:2012, rfc6182}, and improving performance improvement of datacenter networks (DCN)~\cite{Raiciu.et.al:2011, rfc6182} are two impactful applications of MPTCP. While MPTCP is now supported by independent implementations in different operating systems (e.g., Linux, Apple iOS, Citrix, FreeBSD, Oracle Solaris), currently its largest deployment is on mobile devices, likely on hundreds of millions of enabled devices~\cite{rfc8041}.

A MPTCP connection $\mathcal{R}$ consists of a set of one or more subflows $r$. Each subflow provides an alternative path to reach a remote end-system. A subflow starts and terminates as a regular single-path TCP (SPTCP) connection, with its own congestion window, which measures and estimates its state variables, such as RTT. MPTCP packet fields are carried as option into TCP header. Thus, subflow packets are handled as regular TCP packets in the network, which avoids misinterpret and drop of MPTCP packets by middleboxes. This makes easy the deployment of MPTCP on the existing Internet. 

Network paths between multihomed relay/end systems have different characteristics (delay, loss, and capacity), in which MPTCP has to deal with. The reality is that there is a diversity of data traffic and transmission technology in the access networks, i.e., there are several different (heterogeneous) network paths on the Internet. Thus, it is a challenge to improve the performance of multipath transfers to allow a higher bandwidth aggregation benefit. While different algorithms have been proposed to improve multipath performance in different network scenarios and applications in the last years, the research effort has been centered around solutions in the scopes of packet scheduling~\cite{Yang.et.al:2013, Arzani.et.al:2014, Paasch.et.al:2014, Oh.and.Lee:2015, Kimura.et.al:2017}, and coupled congestion controls~\cite{lia:2011, olia:2014, Wischik.et.al:2011, balia:2016, wVegas:2014}. 

In this report, we focus on coupled congestion controls and provide a summay of the existing algorithms deployed in the MPTCP Linux kernel implementation. Simply using standard TCP congestion control on each subflow $r$ would give an unfair resource sharing, since the subflows in an $\mathcal{R}$ connection would dispute bottleneck links with regular SPTCP connections leading them to starvation~\cite{lia:2011}. In addition, the multipath connection undergoes the bottleneck conditions of similar and dissimilar path characteristics. To deal with those concerns, a MPTCP congestion control has to couple and adjust the sending operation of all $r \in \mathcal{R}$ subflows. In doing so, various objectives are taking into account~\cite{Xu.et.al:2016}:
\begin{itemize}
\item \textbf{Friendliness} when sharing network resources with non-MPTCP flows;
\item \textbf{Responsiveness} to react to network changes;
\item \textbf{Throughput improvement} from multipath aggregation bandwidth;
\item \textbf{Congestion balance} among existing multipaths between two end-systems.
\end{itemize}

At present, four coupled congestion controls are deployed in the MPTCP implementation~\cite{Paasch.et.al-mptcp:2017}:  LIA~\cite{lia:2011}, OLIA~\cite{olia:2014}, BALIA~\cite{balia:2016}, and wVegas~\cite{wVegas:2014}. Such congestion controls are mostly carried out in the congestion avoidance phase as there are different strategies for additive-increase/multiplicative-decrease (AIMD), so that the other algorithms (slow start, fast retransmission, and fast recovery) occur as in standard TCP~\cite{rfc5681}. Table~\ref{tab:congestion-controls} summarizes the strategies of the existing multipath congestion avoidance. We provide details of each coupled congestion control in the next sections.

\begin{table*}[]
\caption{Summary of Multipath-Coupled Congestion Controls Deployed in the MPTCP Linux Kernel Implementations.}
\label{tab:congestion-controls}
%\footnotesize
\small
\centering
%\addtolength{\tabcolsep}{0pt}%{-2pt}
\renewcommand{\arraystretch}{1.2}
\setlength\abovedisplayskip{-4pt}
\setlength\belowdisplayskip{0.2pt}

\begin{tabular}[]{ r p{4.4cm}  c  p{5.4cm} c p{4.5cm} }%c }
%\begin{tabular}[]{ r p{4.5cm}  c  p{5.5cm} c p{4.5cm} }%c }
%\multicolumn{7}{ l }{\vspace{-0.2cm}} 
%\\
%\multicolumn{7}{ l }{\textbf{Linked Increases Algorithm (LIA)}~\cite{lia:2011}}
%\\ 
\hline
& \textbf{Alpha Parameter}
& & \textbf{Congestion Window Increase}
& & \textbf{Congestion Window Decrease}
%&
\\ \cline{2 - 2} \cline{4 - 4} \cline{6 - 6}
\textbf{LIA:}
		&		
		\begin{eqnarray}
		\alpha = w_\text{total} ~ \frac{\max \left( {w_r} / {\tau_r^2} \right) }{ \left[ \sum_{k \in \mathcal{R}}{(w_k/\tau_k)} \right] ^2} \nonumber
		\end{eqnarray}
		& &
		\begin{eqnarray}
		\nonumber
		\left. \text{For each ACK received on subflow}~r, \right. \\ \nonumber
		w_r = w_r + \min  \left( \frac{\alpha}{w_\text{total}},  \frac{1}{ w_r } \right)
		\end{eqnarray}				
		& &
		\begin{eqnarray}
		\text{For each packet loss on subflow}~r, \nonumber \\ 
		w_r =  w_r - \frac{w_r}{2}, \nonumber \\ 
		\text{as in standard TCP \cite{rfc5681}} \nonumber		
		\end{eqnarray}
%		&
%		\\
%		\hline
%\multicolumn{7}{ l }{\vspace{-0.2cm}} 
%\\
%\multicolumn{7}{ l }{\textbf{Opportunistic Linked Increases Algorithm (OLIA)}~\cite{olia:2014}}
%\\ \hline
%& \textbf{Alpha Parameter}
%& & \textbf{Congestion Window Increase}
%& & \textbf{Congestion Window Decrease}
%&
\\ \cline{2 - 2} \cline{4 - 4} \cline{6 - 6}
\textbf{OLIA:}
		&
		\begin{eqnarray}
		\nonumber
		 \alpha_r =
						\begin{cases}		
									\frac{1 / |\mathcal{P}|}{|\mathcal{C}|},		& \text{if } r \in \mathcal{C}, \\
									\frac{1 / |\mathcal{P}|}{ |\mathcal{W}|},		& \text{if } r \in \mathcal{W}, |\mathcal{C}| > 0, \\
									0,																					& \text{otherwise}																	
	 					\end{cases}
	 	\end{eqnarray}				
		& &
		\begin{eqnarray}
		\nonumber
		\text{For each ACK received on subflow}~r, \\ \nonumber
		w_r = w_r + \left ( \frac{w_r/\tau_r^2}{\left[ \sum_{k \in \mathcal{R}}{(w_k/\tau_k)} \right] ^2} + \frac{\alpha_r}{w_r} \right)
		\end{eqnarray}
		& &
		\begin{eqnarray}
		\nonumber
		\text{For each packet loss on subflow}~r, \\ \nonumber
		w_r =  w_r - \frac{w_r}{2}, \\ \nonumber		
		\text{as in standard TCP \cite{rfc5681}}
		\end{eqnarray}
%		&
%		\\			
%		\hline
%\multicolumn{7}{ l }{\vspace{-0.2cm}} 
%\\
%\multicolumn{7}{ l }{\textbf{Balanced Linked Adaptation (BALIA)}~\cite{balia:2016}} 
%\\ \hline
%& \textbf{Alpha Parameter}
%& & \textbf{Congestion Window Increase}
%& & \textbf{Congestion Window Decrease}
%&
\\ \cline{2 - 2} \cline{4 - 4} \cline{6 - 6}
\textbf{BALIA:}
		&
		\begin{eqnarray}
			\nonumber
			\alpha_r = \frac{\max\{x\}}{x_r}, \text{ where } x_r = \frac{w_r} {\tau_r}, \\
			\nonumber 
			\text{and } x =  \left[ \sum_{k \in \mathcal{P}}{(w_k/\tau_k)} \right] ^2		
		\end{eqnarray}
		& &
		\begin{eqnarray}
			\nonumber
			\text{For each ACK received on subflow}~r, \\ \nonumber
			w_r =  w_r + \left[ \frac{x_r}{\tau_r ~ x_k} ~ \left(  \frac{1 + \alpha_r}{2} \right ) ~ \left (  \frac{4 + \alpha_r}{5} \right) \right]
		\end{eqnarray}
		& &
		\begin{eqnarray}
			\nonumber
			\text{For each packet loss on subflow}~r, \\ \nonumber
		 	w_r =  w_r - \left[ \frac{w_r}{2} ~ \min \left( \alpha_r, \frac{3}{2}\right) \right]
		\end{eqnarray}
%		&		
%		 \\ 
%		 \hline
%\multicolumn{7}{ l }{\vspace{-0.2cm}} 
%\\
%\multicolumn{7}{ l }{\textbf{Weighted Vegas (wVegas)}~\cite{wVegas:2014}}
%\\ \hline
%& \textbf{Alpha Parameter}
%& & \textbf{Congestion Window Increase}
%& & \textbf{Congestion Window Decrease}
%&
\\ \cline{2 - 2} \cline{4 - 4} \cline{6 - 6}
\textbf{wVegas:}
		&
		\begin{eqnarray}
		\nonumber
		\text{\textbf{if} }  (\delta_r > \alpha_r),  \text{ \textbf{then} } \alpha_r   = \omega_r ~ \alpha_\text{total}, 	\\ \nonumber
		\text{where } \delta_r = \left( \frac{w_r}{\hat{\tau}_r} - \frac{w_r}{\bar{\tau}_r} \right) ~ \hat{\tau}_r, \\ \nonumber
		 \omega_r  = \frac{x_r} { \sum_{k \in \mathcal{R}}{x_k}},  \text{ and } x_i  = \frac{w_i} { \bar{\tau}_i}
		\end{eqnarray}				
		& &
		\begin{eqnarray}
		\nonumber
		\text{For each transmission round on subflow}~r, \\ \nonumber
		w_r =
						\begin{cases}		
									w_r - 1,		&  \text{if } \delta_r > \alpha_r , \\
									w_r + 1,		&  \text{if } \delta_r < \alpha_r \\
	 					\end{cases}
	 					\\ \nonumber
	 	\text{\textbf{if} } (q_r > \bar{\tau}_r - \hat{\tau}_r ), \text{ \textbf{then} } q_r  = \bar{\tau}_r - \hat{\tau}_r \\ \nonumber
	 	\text{\textbf{if} } (\bar{\tau}_r - \hat{\tau}_r \geq 2 q_r), \text{ \textbf{then} } w_r = w_r ~  \frac{\hat{\tau}_r}{2 \bar{\tau}_r}
	 	\end{eqnarray}
		& &
		\begin{eqnarray}
		\nonumber
		\text{For each packet loss on subflow}~r, \\ \nonumber
		w_r =  w_r - \frac{w_r}{2}, \\ \nonumber		
		\text{as in standard TCP \cite{rfc5681}}
		\end{eqnarray}
%		&
		\\ \hline
\end{tabular}
\end{table*}

\section{Linked Increases Algorithm (LIA)}

LIA~\cite{lia:2011} has three main objectives: to {improve throughput} at least as well as an SPTCP connection on the best available path; {do no harm} to other SPTCP connections when sharing common bottlenecks; and {balance congestion} by moving traffic from most congested paths to least congested ones. To this end, when receiving an ACK on subflow $r$, the subflow's congestion window $w_r$ is increased by the minimum between two terms. First term describes a multipath window increase defined by $\alpha / w_\text{total}$, where $\alpha$ is a multipath aggressiveness factor, and $w_\text{total} = \sum_{k \in \mathcal{R}} w_k$. The second term is defined by $1 / w_r$, which represents the window increase of a regular SPTCP connection would get at the same scenario. The aggressiveness factor $\alpha$ is such that the multipath goodput is equal to the SPTCP on the best path.

\section{Opportunistic Linked Increases Algorithm (OLIA)}

While LIA forces a tradeoff between optimal congestion balance and responsiveness, OLIA~\cite{olia:2014} is an alternative that provides simultaneously these both properties. When receiving an ACK on the subflow $r$, the congestion window $w_r$ is increased by adding two terms. The first term is an optimal congestion balancing defined by $({w_r/\tau_r^2})/ {\left[\sum_{k \in \mathcal{R}}{(w_k/\tau_k)} \right] ^2}$, where $\tau_r$ is the round-trip time observed on $r$. The second term is added by ${\alpha_r}/{w_r}$, which guarantees responsiveness to react to changes in the current $w_r$.

Unlike LIA, the parameter $\alpha_r$ is specific to each subflow $r$ and applied to shift traffic among the subflows. To this end, subflows are classified into three different sets: $\mathcal{W}$ -- set of subflows with the largest $w_r$; $\mathcal{B}$ -- set of best subflows with larger estimates of bytes transmitted on $r$ between the last two losses; and $\mathcal{C}$ -- set of the best collected subflows with no larger windows, i.e., $r \in \mathcal{B}$ and $r \notin \mathcal{W}$. If all $r \in \mathcal{B}$ have the largest $w_r$, then $\alpha_r=0$ for any $r$, i.e., the aggregate capacity available is using all the best subflows. 

When $\mathcal{C}$ is not empty, there is at least one best subflow with a small $w_r$, then $\alpha_r$ is positive for all $r \in \mathcal{C}$ and negative for all $r \in \mathcal{W}$. This makes $w_r$ to increase faster for $r \in \mathcal{B}$ and slower for $r \in \mathcal{W}$. As a result, traffic is re-forwarded from subflows fully utilized ($\mathcal{W}$) to paths with free capacity available ($\mathcal{C}$).

\section{Balanced Linked Adaptation (BALIA)}

LIA might be unfriendliness without any benefit to MPTCP when competing with regular SPTCP flows on shared bottlenecks, while OLIA can be unresponsive to changes in network conditions when the paths experience similar RTTs~\cite{Peng.et.al:2016}. From an optimization model designed to guarantee the existence, uniqueness and stability of the network equilibrium, BALIA~\cite{balia:2016, Peng.et.al:2016} is an alternative of multipath congestion control that allows $w_r$ oscillation up to an ideal level to provide good balance between friendliness and responsiveness. For the subflow on the best path or for a MPTCP connection of a single path ($|\mathcal{R}| = 1$), we have $\alpha_r = 1$. This makes both the increment and decrement of $w_r$ to reduce to the same ones of TCP standard \cite{rfc5681}.

\section{Weighted Vegas (wVegas)}

While LIA, OLIA and BALIA react to path congestion upon packet loss, wVegas~\cite{wVegas:2014} is a delay-based congestion control that estimates the link queueing delay $q_r$ to realize path congestion and, then, proactively adapts $w_r$. The control of wVegas adjusts $\alpha_r$ according to the weight $\omega_r$ -- a ratio between current subflow's sending rate $x_r$ and the aggregate multipath sending rate. The $\alpha_r$ is adjusted whenever it is smaller than $\delta_r$, which is a difference between expected and actual $x_r$. 

On the transmission round on $r$, $w_r$ is adapted in two steps. First, similarly to TCP-Vegas \cite{TCP-vegas:1994}, by incrementing or decrementing $w_r$ according to $\delta_r$ and $\alpha_r$. Second, by readjusting $w_r$ according to $q_r$ in order to drain link queues. This is conducted by monitoring the variation of queue delay $q_r$ through the difference between the average observed RTT ($\bar{\tau}_r$) and the minimum measured RTT ($\hat{\tau}_r$). The value of $w_r$ is decreased by a backoff factor of ${\hat{\tau}_r}/{\left( 2\bar{\tau}_r \right)}$ when detecting the link queue variation is higher than $2 q_r$. Such an approach makes subflow $r$ to occupy fewer link buffers, while reducing packet loss and mitigating \textit{bufferbloat}. Thus, wVegas provides a congestion avoidance phase more sensitive to network changes than other congestion controls based exclusivity on the event of packet loss.

\balance

\section{Conclusions}
\label{sec:conclusions}

The adoption of MPTCP is leveraged by its great benefits provided. When extending the TCP, which is implemented at the kernel-space of operating systems, the MPTCP works completely transparent to the user applications running on end-systems. The multipath control data is conveyed as option field in the regular TCP header, which makes it to be easily deployable on the existing Internet infrastructure. The concurrent multipath transfer over subflows established in a connection can improve network resource usage, provide higher throughput and connection failure tolerance to end-systems. However, to obtain such benefits, the subflows have to dispute bottleneck resources  with other non-MPTCP flows. To deal with this concern, MPTCP has to take into account different properties of network congestion, such as fairness and friendliness, in order to share bottleneck resources. In this context, the coupled congestion control is an internal MPTCP mechanism which plays a key role on the multipath performance. In this report, we use a single notation to summarize the four existing algorithms of coupled congestion controls deployed in MPTCP Linux kernel implementation. Since the algorithms operate mostly during the congestion avoidance phase, we focus attention on describing technical details of the multipath AIMD strategies. 

\section*{Acknowledgement}
We thank the support of the research project -- Grant \#2015/18808-0, São Paulo Research Foundation (FAPESP).

\bibliographystyle{IEEEtran}
\bibliography{references}

%\vspace{-1cm}
\begin{IEEEbiography}[{\includegraphics[scale=0.15]{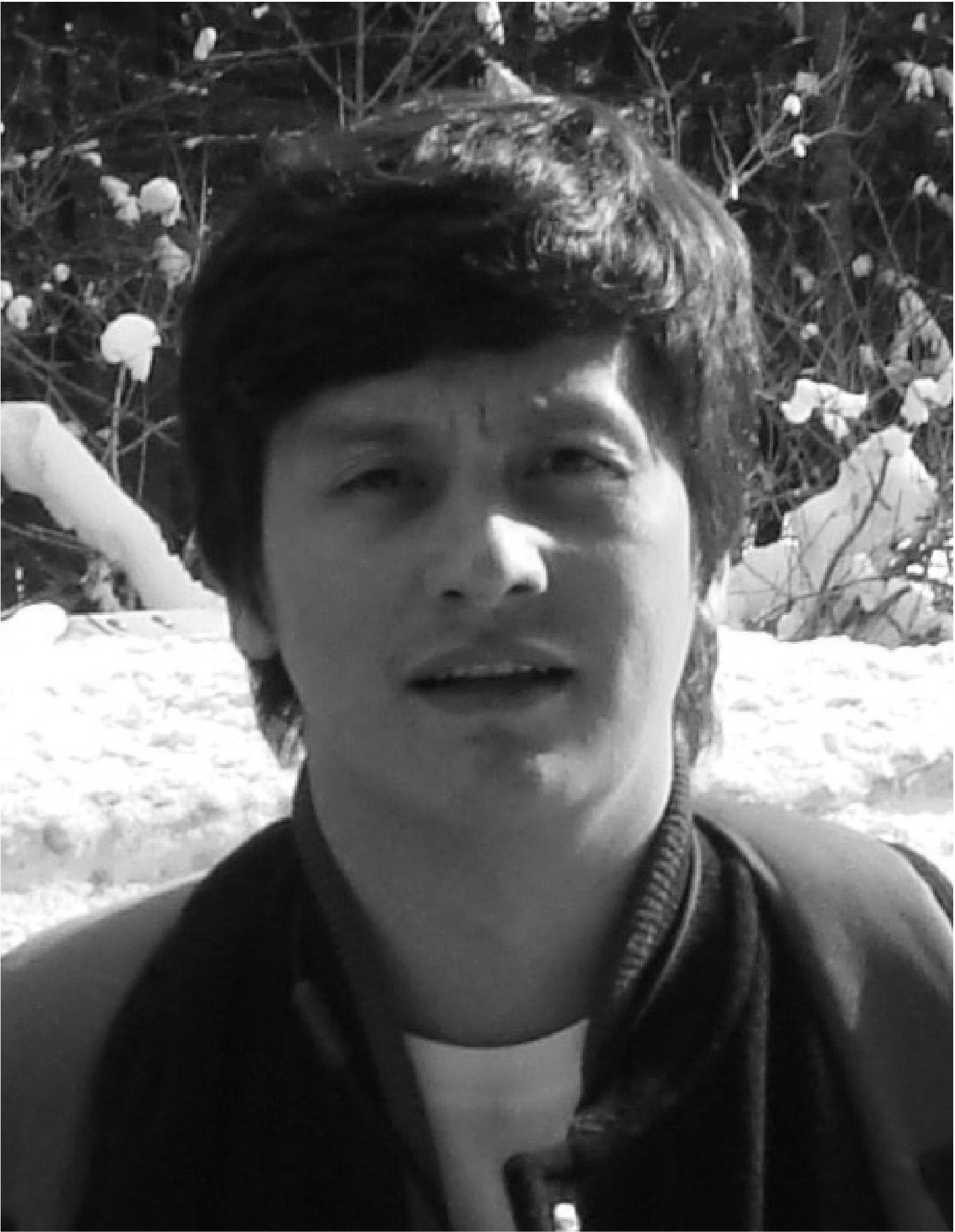}}]{Bruno Yuji Lino Kimura}
received his B.Sc. in Computer Science from the Pontifical Catholic University of Minas Gerais (2005), M.Sc. in Computer Science from the Federal University of São Carlos (2008), and D.Sc in Computer Science from the University of São Paulo (2012), Brazil. Currently, he is an assistant professor at the Federal University of São Paulo. His research interests include multipath data transport, vehicle communication networks, delay/disruption tolerant networks, and mobility management.
\end{IEEEbiography}

%\vspace{-1cm}
\begin{IEEEbiography}[{\includegraphics[scale=0.7]{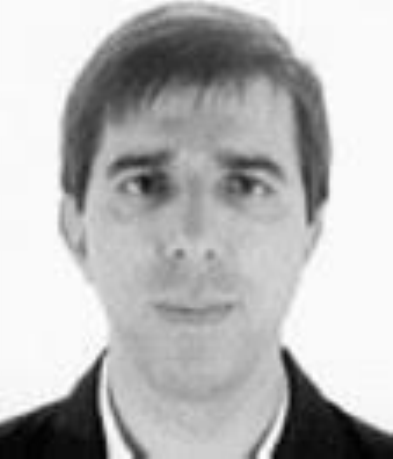}}]{Antonio Alfredo Ferreira Loureiro}
received his B.Sc. and M.Sc. degrees in computer science from the Federal University of Minas Gerais (UFMG), Brazil, and the Ph.D. degree in computer science from the University of British Columbia, Canada. Currently, he is a full professor of computer science at UFMG, where he leads the research group in wireless sensor networks. His main research areas are wireless sensor networks, mobile computing, and distributed algorithms. In the last 10 years he has published regularly in international conferences and journals related to those areas, and also presented tutorials at international conferences.
\end{IEEEbiography}

\end{document}